\documentstyle[aps,twocolumn]{revtex}
\begin{document}
\draft
\input{epsf}

\title{Bell's inequality for the Mach-Zehnder interferometer}

\author{Lars M. Johansen}
\address{Institute of Physics, University of Oslo, P.O.Box 1048
Blindern, N-0316 Oslo, Norway \\ 
Buskerud College, P.O.Box 251, N-3601 Kongsberg, Norway 
\thanks{Permanent address} \thanks{E-mail: lars.johansen@kih.no}}
\date{\today}
\maketitle
\begin{abstract}
We show that no local, hidden variable model can be given for
two-channel states exhibiting both a sufficiently high interference
visibility {\em and} a sufficient degree of anticorrelation
in a Mach-Zehnder interferometer.
\end{abstract}

\section{Introduction}

If a single photon impinges on a Mach-Zehnder interferometer, the
probability of detecting the photon in a certain output channel
depends on the difference between the phase delays imposed by the two
interfering channels. The photon behaves as if it goes through both
interfering channels. On the other hand, if detectors are inserted
into each interfering channel, the photon is only detected in one of
the channels. In the delayed-choice experiment, the detectors may
even be inserted in the last instant. Then the experimenter's choice
whether to insert these detectors or not determines whether the
photon shall behave ``as if" it goes through one or both interfering
channels.

It might seem that the detectors impart some sort of ``nonlocal
collapse" to the photon state. Otherwise, one may imagine that the
photon itself goes through one channel only accompanied by an ``empty
wave" in the other channel. Such models have been developed by de
Broglie \cite{deBroglie27} and later by Bohm \cite{Bohm}. They are
known to be nonlocal.

A common feature of both the ``collapse" picture and the ``empty
wave" picture is that they refer to systems which appear to be
``delocalized" (yielding interference) in a certain experiment and
``localized" (yielding anticorrelation) in another. It may seem that
such behavior is in some way nonlocal. 

Although classical optical fields may display interference effects in
the same way as single photons, they cannot simultaneously be
anticorrelated. One may also imagine stochastic classical fields
which yield anticorrelation between two channels, but such
two-channel systems may not give rise to interference effects when
superposed.

One might suspect that two-channel quantum states which yield {\em
both} interference {\em and} anticorrelation in a Mach-Zehnder
interferometer violate local realism. In this paper we shall see that
this is indeed the case, and that for such systems no local hidden
variable model \cite{Bell64} can be constructed.

\section{The Mach-Zehnder interferometer}

\begin{figure}
\leavevmode
\centering
\epsfxsize=2.5in
\epsfbox{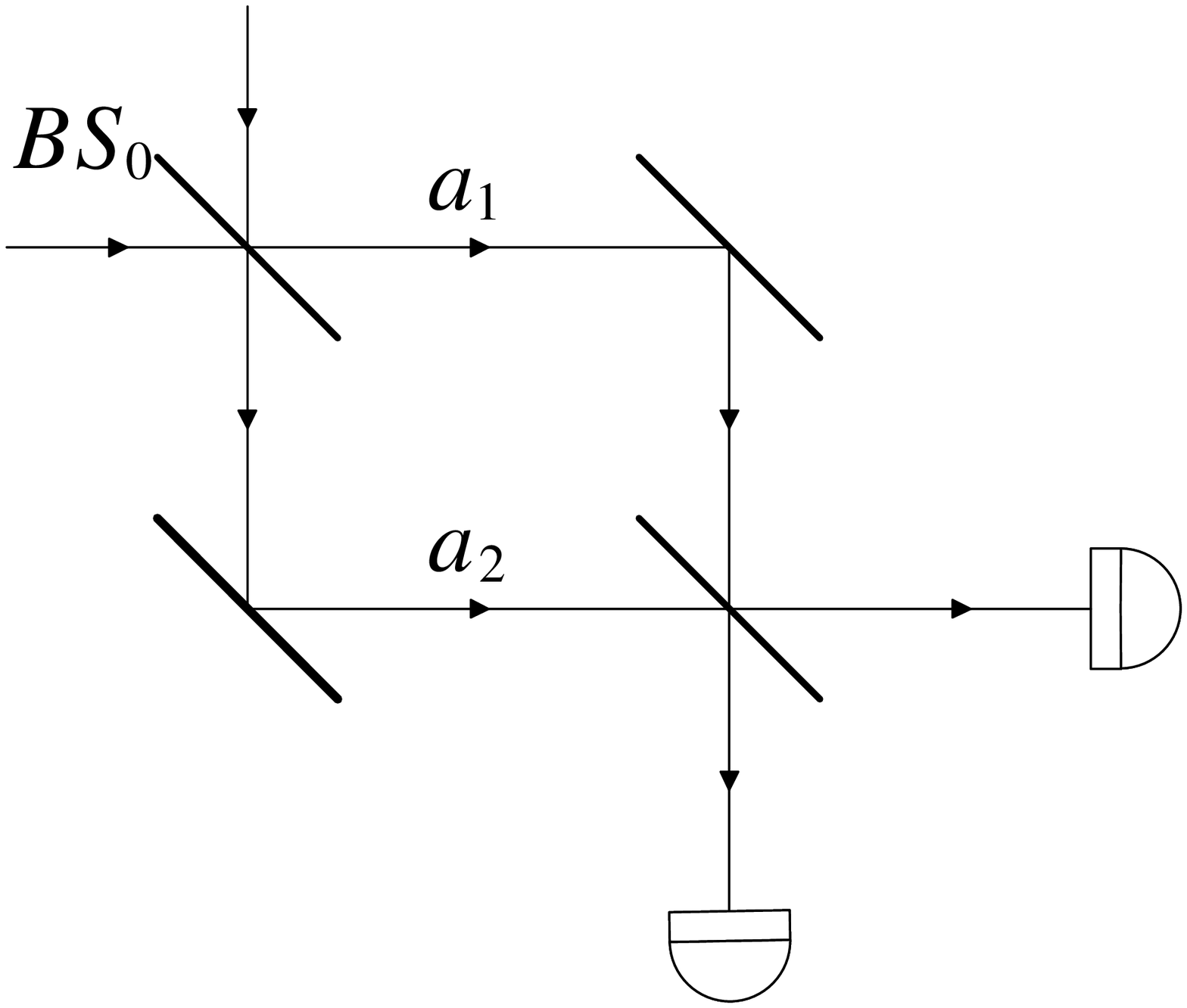}
	\caption{Mach-Zehnder interferometer where the degree of first
	and second order coherence between the $a$-channels may be
	observed directly.} 
	\label{fig:machzehnder}
\end{figure}

Before deriving the Bell inequality, we first need to specify the
characteristic quantities that we observe in a Mach-Zehnder
interferometer (see Fig. \ref{fig:machzehnder}).

It is well known that the interference visibility obtainable in a
first
order interferometer is represented by the degree of first order
coherence \cite{Glauber63a}
\begin{equation}
    g^{(1)} = {\langle \hat{a}_1^{\dag} \hat{a}_2
    \rangle \over \sqrt{\langle \hat{a}_1^{\dag} \hat{a}_1 \rangle
    \langle \hat{a}_2^{\dag} \hat{a}_2 \rangle}}.
\end{equation}
Here $\hat{a}_k$ and $\hat{a}_k^{\dag}$ are annihilation and creation 
operators for the interfering channel $a_k$ ($k=1,2$) of the
interferometer. For simplicity we employ a single mode treatment.

If, on the other hand, we insert detectors into each channel of the
interferometer, we may observe the coincidence rate or the degree of
second order coherence between the same two channels
\cite{Glauber63a},
\begin{equation}
    g^{(2)} = {\langle \hat{a}_1^{\dag} \hat{a}_2^{\dag}
    \hat{a}_2 \hat{a}_1 \rangle \over \langle
    \hat{a}_1^{\dag} \hat{a}_1 \rangle \langle \hat{a}_2^{\dag}
    \hat{a}_2 \rangle}.
\end{equation}
The fact that classical field theories do not allow both interference
and anticorrelation for the same system is illustrated by the fact
that in these theories the inequality \cite{TitulaerGlauber65}
\begin{equation}
   \mid g^{(1)} \mid^2 \le g^{(2)}
   \label{eq:titulaerglauber}
\end{equation}
must be fulfilled.

\section{Bell's inequality}

\begin{figure}
\leavevmode
\centering
\epsfxsize=3.0in
\epsfbox{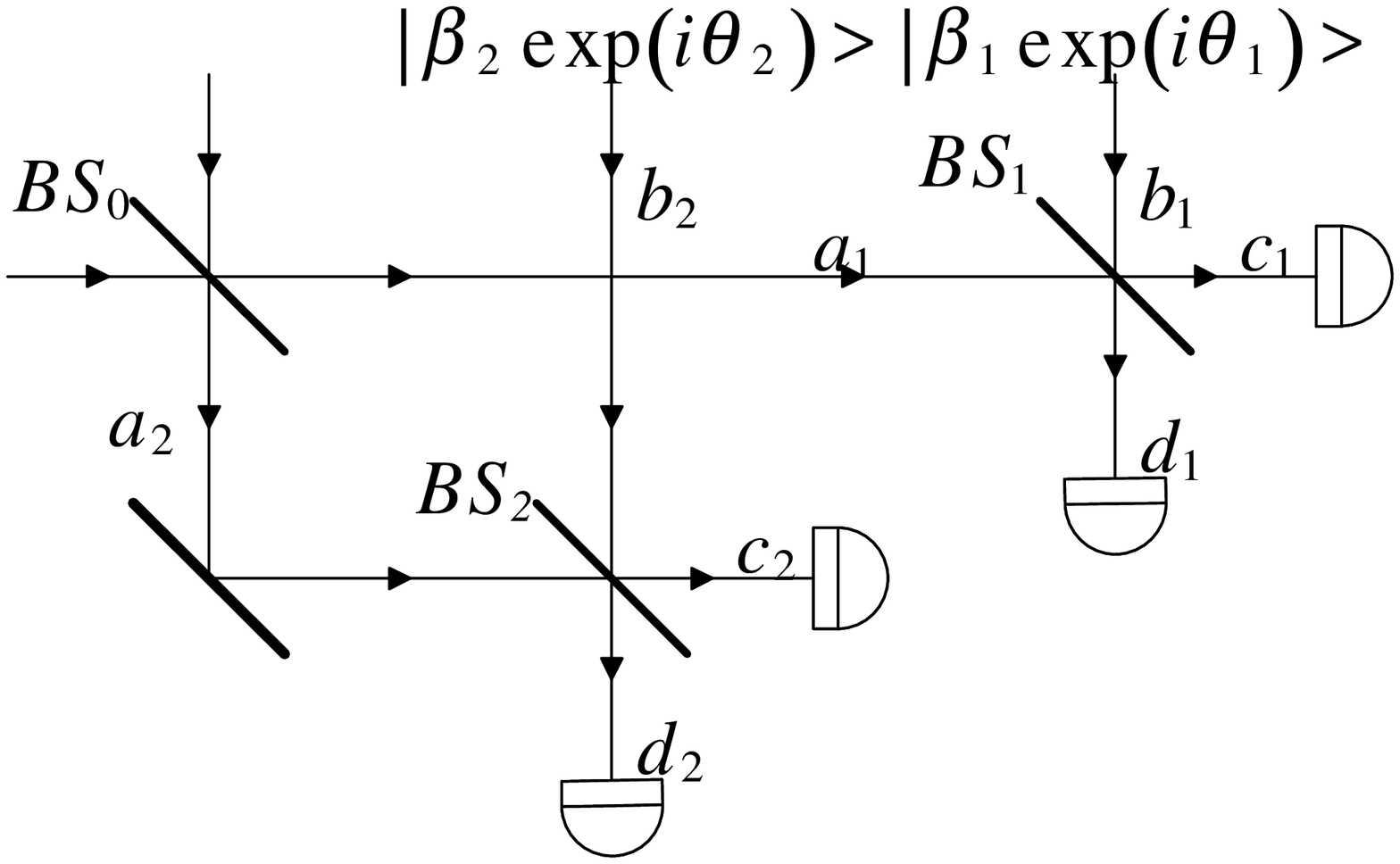}
	\caption{An experiment where the two $a$-channels are mixed with
	with coherent, homodyne local oscillators. In this experiment,
	local realism imposes constraints on the degree of first and
	second order coherence between the $a$-channels.}
	\label{fig:bell}
\end{figure}

We first analyze the experiment shown in Fig. \ref{fig:bell}. We
shall see that the observables here are closely related to the
Mach-Zehnder interferometer (Fig. \ref{fig:machzehnder}).

Grangier {\em et al.} \cite{GrangierPotasekYurke88} were the
first to propose the use of local oscillators in Bell experiments.
The Bell experiment that we will use here (Fig. \ref{fig:bell}) was
first proposed by Oliver and Stroud \cite{OliverStroud89}. They also
showed that single photon states violate local realism in this
interferometer. Later Tan, Holland and Walls (THW)
\cite{TanHollandWalls90} performed a thorough derivation of the
conditions for local realism for any state in this interferometer.
The behavior of single photon states in this interferometer was
treated extensively by Tan, Walls and Collett
\cite{TanWallsCollett91}.

We shall generalize the work of THW. Again, we restrict the attention
to single-mode systems. Let $\hat{\mu}_k$ and $\hat{\mu}_k^{\dag}$ be
annihilation and creation operators for the channel $\mu_k$
($\mu=a,b,c,...$, $k=1,2$). The $a$-channels are output channels from
beamsplitter $BS_0$, in analogy with the Mach-Zehnder interferometer
(Fig. \ref{fig:machzehnder}). Each channel $a_k$ is mixed with a
local oscillator channel $b_k$ on a semireflecting beamsplitter
$BS_k$. The local oscillator is represented as a coherent state $\mid
\beta_k \exp(i \theta_k)\rangle$ with (real) amplitude $\beta_k$ and
(real) phase $\theta_k$.

The connection between the input and output on beamsplitter $BS_k$ is
given by the transformation
\begin{equation}
 \left ( \begin{array}{c}  \hat{c}_k \\ \hat{d}_k \end{array} \right
 ) =  {1 \over \sqrt{2}} \left ( \begin{array}{cc} 1 & i \\ i & 1
 \end{array} \right ) \left ( \begin{array}{c}  \hat{a}_k \\
 \hat{b}_k \end{array} \right ).
\end{equation}
We find that the total photon number is
preserved,
\begin{mathletters}
\begin{equation}
	\hat{S}_k = \hat{c}_k^{\dag} \hat{c}_k + \hat{d}_k^{\dag}
	\hat{d}_k = \hat{a}_k^{\dag} \hat{a}_k + \hat{b}_k^{\dag}
	\hat{b}_k.
\end{equation}
Also, we may rewrite the difference between photon numbers at the two 
beamsplitter output channels in terms of input operators,
\begin{equation}
	\hat{D}_k = \hat{c}_k^{\dag} \hat{c}_k - \hat{d}_k^{\dag}
	\hat{d}_k = i \left ( \hat{a}_k^{\dag} \hat{b}_k - \hat{a}_k
	\hat{b}_k^{\dag} \right ).
\end{equation}
    \label{eq:sumdiff}
\end{mathletters}
We will now study the quantity
\begin{equation}
    E(\theta_1,\theta_2) = {\langle \hat{D}_1 \hat{D}_2 \rangle \over
    \langle \hat{S}_1 \hat{S}_2 \rangle},
\end{equation}
which may be termed the ``modulation depth" of the correlation
between the two interferometers. Note that it's modulus is restricted
from above to unity.

Using the operator definitions (\ref{eq:sumdiff}) we find
\begin{eqnarray}
 E(\theta_1,\theta_2) = {\beta_1 \beta_2 \over \langle
 \hat{a}_1^{\dag} \hat{a}_1 \hat{a}_2^{\dag} \hat{a}_2 \rangle +
 \langle \hat{a}_1^{\dag} \hat{a}_1 \rangle \beta_2^2 + \langle
 \hat{a}_2^{\dag} \hat{a}_2 \rangle \beta_1^2 + \beta_1^2 \beta_2^2 }
 \nonumber \\  \times  \left [ \langle \hat{a}_1^{\dag} \hat{a}_2
 \rangle \exp[i(\theta_1-\theta_2)] + \langle \hat{a}_1
 \hat{a}_2^{\dag} \rangle \exp[-i(\theta_1-\theta_2)] \right . \nonumber
 \\ - \left . \langle \hat{a}_1^{\dag} \hat{a}_2^{\dag} \rangle
 \exp[i(\theta_1+\theta_2] - \langle \hat{a}_1 \hat{a}_2 \rangle
 \exp[- i(\theta_1+\theta_2)] \right ]. \label{eq:bigbad}
\end{eqnarray}
The local oscillator amplitudes are of course independent parameters,
and may be chosen freely. We now want to choose them so that the
modulation depth is maximized. It can be shown that this is achieved
by the choice
\begin{equation}
 \beta_1 \beta_2 = \sqrt{\langle  \hat{a}_1^{\dag} \hat{a}_1
 \hat{a}_2^{\dag} \hat{a}_2 \rangle}, \,\,\,\,  {\beta_1 \over
 \beta_2} = \sqrt{{\langle \hat{a}_1^{\dag} \hat{a}_1 \rangle \over
 \langle \hat{a}_2^{\dag} \hat{a}_2 \rangle }}.
\end{equation}
Inserting this into Eq. (\ref{eq:bigbad}) we may write
\begin{eqnarray}
	E(\theta_1, \theta_2) &=& C_1 \cos(\theta_1-\theta_2+ \arg 
	\langle \hat{a}_1^{\dag} \hat{a}_2 \rangle) \nonumber \\ &+&
	C_2 \cos(\theta_1+\theta_2 + \arg  \langle \hat{a}_1^{\dag}
	\hat{a}_2^{\dag} \rangle ).
 \label{eq:trigform}
\end{eqnarray}
We are particularly interested in the coefficient $C_1$, since it may
be expressed in terms of well known coherence functions between the
two $a$-channels,
\begin{equation}
 C_1 = { \mid g^{(1)} \mid \over 1 + \sqrt{g^{(2)}}}.
\end{equation}
If we consider the quantity
\begin{equation}
    B = E(\theta_1,\theta_2) + E(\theta_1,\theta_2') +
    E(\theta_1',\theta_2) - E(\theta_1',\theta_2'),
\end{equation}
local realism imposes the restriction \cite{Assumption}
\begin{equation}
	\mid B \mid \le 2.
	\label{eq:CHSH}
\end{equation}
THW \cite{TanHollandWalls90} showed that this is satisfied whenever
\begin{equation}
	C_1^2 + C_2^2 \le 1/2.
\end{equation}
Therefore a minimal (necessary, but not sufficient) requirement for
local realism is that
\begin{equation}
    C_1 \le 1/\sqrt{2},
    \label{eq:localrealism}    
\end{equation}
or
\begin{equation}
 { \mid g^{(1)} \mid \over 1 + \sqrt{g^{(2)}}} \le 1/\sqrt{2}.
    \label{eq:bell}
\end{equation}
This may be considered as a rewriting of the Bell inequality
(\ref{eq:CHSH}). We see that violation of this inequality implies
that inequality (\ref{eq:titulaerglauber}) is also violated. However,
inequality (\ref{eq:titulaerglauber}) must be strongly violated in
order to imply violation of inequality (\ref{eq:bell}). In other
words, if the state violates local realism, it also violates
classical field theory, but the converse is not necessarily true.

\section{Bell's inequality for the Mach-Zehnder interferometer}

We of course note that the parameters involved in Bell's inequality
(\ref{eq:bell})  are exactly the same that we observe in the
Mach-Zehnder interferometer, namely the degree of first and second
order coherence. In other words, this inequality involves the
interference visibility and the coincidence rate for a Mach-Zehnder
interferometer. This of course also means that we can {\em test} this
inequality in a Mach-Zehnder interferometer.

It can be noted that in order to observe violation of inequality
(\ref{eq:bell}), a minimal requirement is that
\begin{eqnarray}
    g^{(1)} &>& 1/\sqrt{2} \approx 0.71, \nonumber \\
    g^{(2)} &<& (\sqrt{2}-1)^2 \approx 0.17.
\end{eqnarray}
We see that the state must be both sufficiently anticorrelated and it
must yield a sufficiently high interference visibility. We see that
this corresponds with our suspicion that the combination of these two
features in some way yields violation of local realism.

The most extreme example in this respect is of course the split
single photon state
\begin{equation}
    \mid \psi \rangle_a = {1 \over \sqrt{2}} \left ( \mid 1
    \rangle_{a_1} \mid 0 \rangle_{a_2} + i \mid 0 \rangle_{a_1}
    \mid 1 \rangle_{a_2} \right ),
    \label{eq:singlephoton}
\end{equation}
which yields $g^{(1)} = 1$ and $g^{(2)} = 0$. This state has been
shown to violate local realism
\cite{OliverStroud89,TanHollandWalls90,TanWallsCollett91} in the same
experiment as we consider here. But in this paper we have in addition
seen that a whole class of states violates local realism, including
also mixed states. These states possess certain common features,
namely those of a high interference visibility in combination with a
strong anticorrelation.

Grangier, Roger and Aspect \cite{GrangierRogerAspect86} have
performed an experiment measuring the visibility and the coincidence
rate in the Mach-Zehnder interferometer. They observed a visibility
of 0.98 and a coincidence rate of 0.18. Although this is sufficient
to demonstrate violation of inequality (\ref{eq:titulaerglauber}),
the coincidence rate was slightly too high to demonstrate violation
of inequality (\ref{eq:bell}). However, such a demonstration should
be well within technological reach today.

Note that a direct contradiction with local realism is not achieved
in a Mach-Zehnder interferometer. If inequality
(\ref{eq:titulaerglauber}) is violated, the experiments do show
that classical field theories break down. However, even if inequality
(\ref{eq:bell}) is violated, the results can be explained by a {\em
local} but {\em contextual} hidden variable model. We may, e.g.,
explain the interference visibility in terms of a classical wave
model and the anticorrelation in terms of a classical particle model.

Still, it is interesting to see that merely by observing the
interference visibility and the coincidence rate on an unknown state,
we may gain sufficient information to predict that this state will
violate local realism in the THW-experiment. Thus, any single-mode
state, pure or mixed, which displays both a sufficiently high
interference visibility and a sufficient degree of anticorrelation
violates local realism.

\section*{Acknowledgments}

The author wishes to thank Kristoffer Gj{\o}tterud, Paul Kwiat, Barry
Sanders and Aephraim Steinberg for useful discussions and comments on
an earlier version of this paper. This work was financed by the
University of Oslo, and is a cooperative project with Buskerud
College.


\begin{references}

\bibitem{deBroglie27}
L. {de Broglie}, J. Phys. Radium, Series 6, 5 (1927) 225.

\bibitem{Bohm}
D. Bohm, Phys. Rev. 85 (1952) 166, 180.

\bibitem{Bell64}
J.~S. Bell, Physics (Long Island City, N.Y.) 1 (1964) 195.

\bibitem{Glauber63a}
R.~J. Glauber, Phys. Rev. 130 (1963) 2529.

\bibitem{TitulaerGlauber65}
U.~M. Titulaer and R.~J. Glauber, Phys. Rev. 140 (1965) B676.

\bibitem{GrangierPotasekYurke88}
P. Grangier, M.~J. Potasek, and B. Yurke, Phys. Rev. A 38 (1988)
3132.

\bibitem{OliverStroud89}
B.~J. Oliver and C.~R. {Stroud Jr.}, Phys. Lett. A 135 (1989) 407.

\bibitem{TanHollandWalls90}
S.~M. Tan, M.~J. Holland, and D.~F. Walls, Opt. Commun. 77 (1990)
285.

\bibitem{TanWallsCollett91}
S.~M. Tan, D.~F. Walls, and M.~J. Collett, Phys. Rev. Lett. 66 (1991)
252.

\bibitem{Assumption} See, e.g., the development in D.~F. Walls and
G.~J. Milburn, Quantum Optics (Springer-Verlag, Berlin, 1994).
It should be noted that this derivation involves an additional
``no-enhancement" assumption (J.~F. Clauser and A. Shimony, Rep.
Prog. Phys. 41 (1978) 1881). This is a common feature of all Bell
inequalities that have been experimentally tested so far.
Inequalities derived without this assumption generally require higher
detector efficiencies in order to be tested.

\bibitem{GrangierRogerAspect86}
P. Grangier, G. Roger, and A. Aspect, Europhys. Lett. 1 (1986) 173.

\end{references}
\end{document}